\documentclass[aps,prl,preprint,amsfonts,superscriptaddress,showpacs,letterpaper]{revtex4}
\usepackage{graphicx}
\usepackage{bm}

\begin{document}

\title{Coarse-graining protein energetics in sequence variables}

\author{Fei Zhou}
\affiliation{Department of Physics,
 Massachusetts Institute of Technology,
 Cambridge, MA 02139, USA}
\author{Gevorg Grigoryan}
\affiliation{Department of Biology,
 Massachusetts Institute of Technology,
 Cambridge, MA 02139, USA}
\author{Steve R.\ Lustig}
\affiliation{DuPont Central Research and Development,Experimental Station, Wilmington, Delaware 19880, USA}
\author{Amy E.\ Keating}
\affiliation{Department of Biology,
 Massachusetts Institute of Technology,
 Cambridge, MA 02139, USA}
\author{Gerbrand Ceder}
\affiliation{Department of Materials Science and Engineering,
 Massachusetts Institute of Technology,
 Cambridge, MA 02139, USA}
\author{Dane Morgan}
\affiliation{Department of Materials Science and Engineering,
 Massachusetts Institute of Technology,
 Cambridge, MA 02139, USA}

\date{\today}

\begin{abstract}
We show that cluster expansion (CE), previously used to model solid-state materials with
binary or ternary configurational disorder, can be extended to the protein design problem. We present a generalized CE framework suitable for protein
studies, in which properties such as the energy can be
unambiguously expanded in the amino-acid sequence space. The CE
coarse-grains over non-sequence degrees of freedom (e.g.,
sidechain conformations) and thereby simplifies the problem of
designing proteins, or predicting the compatibility of a sequence
with a given structure, by many orders of magnitude. The CE is
physically transparent, and can be evaluated through linear
regression on the energy of training sequences. The approach is demonstrated on two distinct backbone folds. We show that good prediction
accuracy is obtained with up to pairwise interactions for a
coiled-coil backbone, and that triplet interactions are important
in the energetics of a more globular zinc-finger backbone.

\end{abstract}

\maketitle

Protein folding and protein design stand among the most formidable
challenges in contemporary computational biology. The 3-D structure of a protein is uniquely encoded in its 1-D sequence of amino acids (AA), and enormous theoretical and computational research effort has been devoted to understanding how
\cite{Kraemer-Pecore2001Computational,Pokala2001Review:,Vendruscolo2003Protein}.
The problem can be posed two ways: protein {\it folding} deals with predicting the
final 3-D structure of a protein given its
AA sequence, whereas protein {\it design} is concerned with finding
an optimal sequence to fold to a pre-defined structure. Protein
design is useful both because it allows for the engineering of
macromolecules with desired properties
\cite{Dwyer2004Computational,Allert2004Computational,Kuhlman2003Design},
and because the development of computational design methods
deepens our general understanding of protein folding and
stability. Scoring functions that indicate the ability of
sequences to fold to any given structure are central to both the folding and design problems. These range from statistical
knowledge-based functions derived from databases of known protein
structures \cite{Russ2002Knowledge-based} to empirical functions
mainly based on experimental measurements
\cite{Guerois2001Protein}, to more physics-based functions that
attempt to model protein free energy
\cite{Guerois2001Protein,Mendes2002Energy}.

\begin{figure}[tph]
\includegraphics[width=0.99 \linewidth]{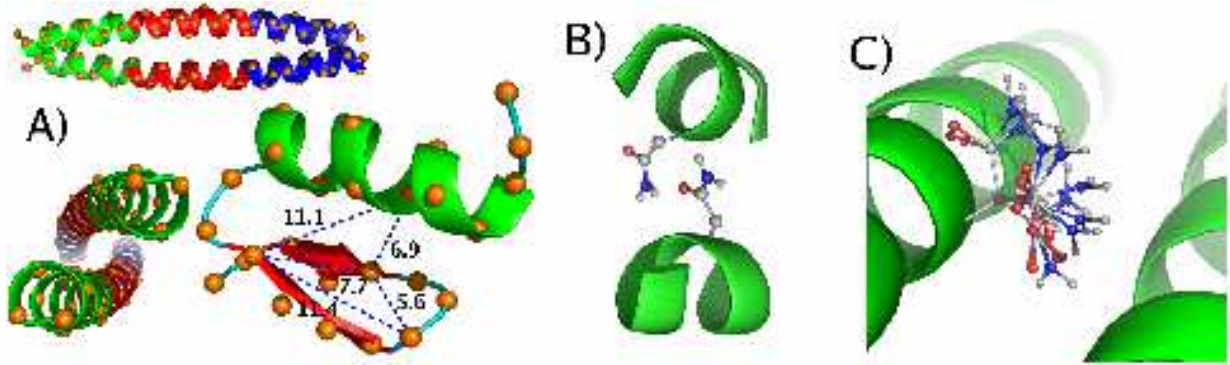}
\caption{A) Two protein folds - the coiled
coil (top - side view; left - helix axis view) and the zinc-finger (right).
Orange spheres are backbone atoms and the ribbons are a cartoon
representation of the backbone geometry. The coiled-coil unit cells are
highlighted. B) The optimal
rotamers for two AA's in an all-atom representation. C) A
set of common rotamers for one AA shown superimposed.}
\label{fig:peptide-structure}
\end{figure}

Physics-based energy
functions have the potential of being the most accurate and
interpretable. These express the energy of a protein sequence adopting a specified structure in terms of atomic coordinates, and account for energies arising from van der Waals (vdW) forces, electrostatics, and solvation. All atoms in a protein can be classified as either ``backbone'' or ``side-chain''. The backbone atoms are the same for each AA and represent the overall structure or ``fold'' of a protein, as shown for two examples in Fig.\ \ref{fig:peptide-structure}. The side-chain atoms are different for different AAs, and give rise to additional degrees of freedom termed ``side-chain conformations'' or ``rotamers'' (see Fig.\ \ref{fig:peptide-structure}B-C).  Even for a relatively small protein
fold of 100 AAs there are roughly  $10^{130}$ possible
sequences. Accounting for side-chain conformations expands the search space to $\tilde 10^{230}$ structures. The computational complexity of high-quality physics-based scoring functions makes a search for optimal sequences intractable. Because sequence determines the structure of a protein, however, a function should exist that maps sequence directly to energy. A sufficiently accurate and computationally tractable approximation of this function would find wide spread use in computational studies of protein structure.

Mapping sequence to energy is similar to the configurational problem in alloy theory
\cite{Sanchez1984Generalized,de1994,Ceder1993derivation}
 where distributions of A and B atoms on a fixed topology
of lattice sites specifies the energy
\cite{Garbulsky1995Linear-Programming,Blum2004Structural,
Wolverton2002Incorporating,Ceder1997model}. The technique of cluster expansion (CE) \cite{Sanchez1984Generalized,de1994} has proven extremely useful for rapidly evaluating the energies of alloys and searching for low-energy configurations. In this Letter, we apply CE to the protein design problem, deriving two structure-specific functions that can determine the energies of a sequence adopting either a coiled-coil or a zinc-finger geometry. Searches using these functions can be used in the future to identify low-energy sequences that adopt these folds. Further, CE can potentially be applied directly to the more challenging protein folding problem by deriving a function specific to each of the $~\sim$1,000 known protein folds. Rapid evaluation of a sequence with the full panel of functions could identify the best structure. This approach, termed ``threading'' or ``fold recognition'', is widely used for structure prediction in combination with statistically derived energy functions.

While in alloys one typically treats binary distributions (two possible species per site) or on rare occasions ternaries \cite{Althoff1995Commensurate,McCormack1995Nonempirical}, the general protein design problem requires extension to all twenty possible AAs.
For a protein of $L$ residues let the variable $\sigma^i =1 \dots m$ indicate which of the $m$
AAs is present at site $i$. A sequence is then expressed by $\vec\sigma = \{
\sigma^1, \dots ,  \sigma^L \}$.  The energy of a protein $E[\vec\sigma,\vec{\tau}]$ depends on this sequence and on the other
microscopic information $\vec\tau$ (e.g.\ positions of all atoms on the protein and
solvent molecules).
The important energy function in protein design, $E_{\min}[\vec\sigma]$, can be obtained by optimizing over $\vec\tau$:
\begin{equation}
E_{\min}[\vec\sigma] = \min_{\vec{\tau}} E[\vec\sigma,\vec{\tau}]. \label{eq:coarsegrain}
\end{equation}
The CE is a general approach to obtain
$E_{\min}$ by expanding in a suitable set of independent basis
functions.
Let $i, j, k
= 1 \dots  L$ denote AA sites and $\alpha, \beta,\gamma=1 \dots m-1$
index basis functions  $\{1, \phi^i_\alpha,\phi^i_\alpha \phi^j_\beta,\phi^i_\alpha \phi^j_\beta \phi^k_\gamma, \dots\}$. The energy can be expanded as:
%
%
\begin{eqnarray}
E_{\min}[\vec\sigma] &=& J_\emptyset + \sum_{i,\alpha} J^i_\alpha
\phi^i_\alpha(\sigma^i) + \sum_{ij,\alpha \beta} J^{ij}_{\alpha \beta} \phi^i_\alpha(\sigma^i)
\phi^j_\beta(\sigma^j) \nonumber \\
&+& \sum_{ijk,\alpha \beta \gamma} J^{ijk}_{\alpha \beta \gamma}
\phi^i_\alpha(\sigma^i) \phi^j_\beta(\sigma^j)
\phi^k_{\gamma}(\sigma^k) + \dots  \label{eq:CE}
\end{eqnarray}
where the $J$s are expansion coefficients. We leave it to a future paper to describe the mathematical properties of this basis set and to show its completeness in the space of all possible $L$-site AA sequences.
Eqn.\ \ref{eq:CE} is in principle exact, though in
practice the expansion has to be truncated. While the $J$ coefficients depend on the choice of basis functions, the sum over terms
spanning an cluster of AA sites $\{i, \dots , j \}$ has a physical interpretation, and can be defined as effective
interaction (EI)  between the AA's on these sites:
\begin{equation}
EI(\sigma^i \dots \sigma^j)= \sum_{\alpha \dots \beta} J^{i \dots
j}_{\alpha \dots \beta} \phi^{i}_{\alpha}(\sigma^i) \dots
\phi^j_\beta(\sigma^j). \label{eq:EI}
\end{equation}
The choice of point basis functions
$\phi_\alpha$ is in principle arbitrary though we have found that
previously proposed basis functions \cite{Sanchez1984Generalized}
have poor numerical stability for the high dimensional configuration spaces of proteins  and make the expansion converge slowly.
In this Letter we use $\phi_\alpha(\sigma)=\delta(\sigma-\alpha)$.
Hence $\phi_\alpha(m)\equiv 0$ and the hypothetical sequence
$\{m,\dots ,m\}$ has energy $J_0$. If we assign $m$ to
Alanine (Ala) any point EI($\sigma^i$) equals the energetic contribution
of $\sigma^i$ relative to Ala. Therefore, point EIs exactly
correspond to the change in energy upon mutating a residue to alanine, a quantity that is frequently measured experimentally to assess the importance of a residue to stability. Pair EI($\sigma^i,\sigma^j$) is the
interaction of an AA pair. This is also a measure well known to
biochemists \cite{Krylov1994Thermodynamic,Acharya2002heterodimerizing}. This
concept can be taken beyond pairs -- contributions purely from
triplets can be measured similarly. Although this is
difficult to do experimentally, the CE allows one to
systematically analyze the importance of higher order
interactions.

Given $E_{\min}$ for enough sequences, $J$s can be
extracted by standard fitting procedures. Determining which $J$s
to keep in the fit is not always obvious. While one may be guided
by the idea that point terms are larger than pairs, which in turn
are larger than triplets, this is not always true.
We use a more systematic way for evaluating important $J$s based
on the cross-validation (CV) score \cite{van2002Automating}. Essentially, the CV score is the average
error with which each sequence is predicted when left out of the
fitting, and as such is a good measure of the prediction power. Our procedure consists of fitting a selected set of candidate clusters and order them by the
average $|J|$. Clusters for which the $J$ value largely arises
from numeric noise increase the CV score, and are excluded. When a cluster is included, so are all of its sub-clusters.

We demonstrate the power of the CE by testing it on two different protein folds, mimicking the protein design problem.  The folding energy is
defined as the energy difference between the folded and the
unfolded states: $E_{\rm {folding}}= E_{\rm {folded}} - E_{\rm {unfolded}}$.
Although the CE can in principle be used with any energy model, we
test it here with a physically meaningful but relatively simple
expression similar to Hamiltonians commonly used in the design
field \cite{Ali2005Design}:
\begin{eqnarray}
E[\vec{\sigma}, \vec{\tau}] &=& E^{\rm {vdW}} + E^{\rm {elec,wat}} + E^{\rm {solv,sc}} +
E^{\rm {torsion}}, \label{eq:efunction}
\end{eqnarray}
where $E^{\rm{vdW}}$ is the vdW interaction
modeled as a 6-12 Lennard-Jones potential, $E^{\rm {elec,wat}}$ is
the total electrostatic energy (excluding intra-sidechain
interactions), $E^{\rm {solv,sc}}$ is the solvation
energy of all backbone and sidechain atoms
\cite{Lazaridis1999Effective}, and $E^{\rm {torsion}}$ is the
sidechain torsional energy. All energy terms are calculated using
the CHARMM package \cite{Brooks1983Charmm} with the {\it param19}
parameters. The unfolded state is modeled by retaining only
sidechain self energies and local interactions between sidechains
and their surrounding penta-peptide backbone. Because
$E[\vec\sigma, \vec{\tau}]$ in Eq.\ \ref{eq:efunction} is
pairwise-decomposable, we are able to apply the dead-end-elimination
(DEE) algorithm \cite{Desmet1992Dead-End,Goldstein1994Efficient} followed by
a branch-and-bound search to arrive at the optimal sidechain
conformations corresponding to $E_{\min}$. Thus in a CE derived from these $E_{\min}$, the $J$s, and
hence EIs, parameterize optimized energies whereby all the
sidechain degrees of freedom are coarse-grained out.  The EI,
defined at the sequence level, may include higher order terms even
though the initial energy expressions at the conformational level
are pairwise decomposable. The advantage of this procedure is an
enormous reduction in the search space, from $(20m)^L$ to $m^L$,
where 20 is the average number of rotamers considered per AA.

In order to more accurately fit the important low energies, our fitting is weighted by $\max(e^{-(E-E_0)/K}, w_0)$, where
$E_0$ is the lowest energy in the data set, $K$ is
approximately the range of interest above $E_0$ and $w_0$ is the minimal weight
at large $E$ to avoid numeric instability.

\begin{figure}[tph]
\includegraphics[width=0.8 \linewidth]{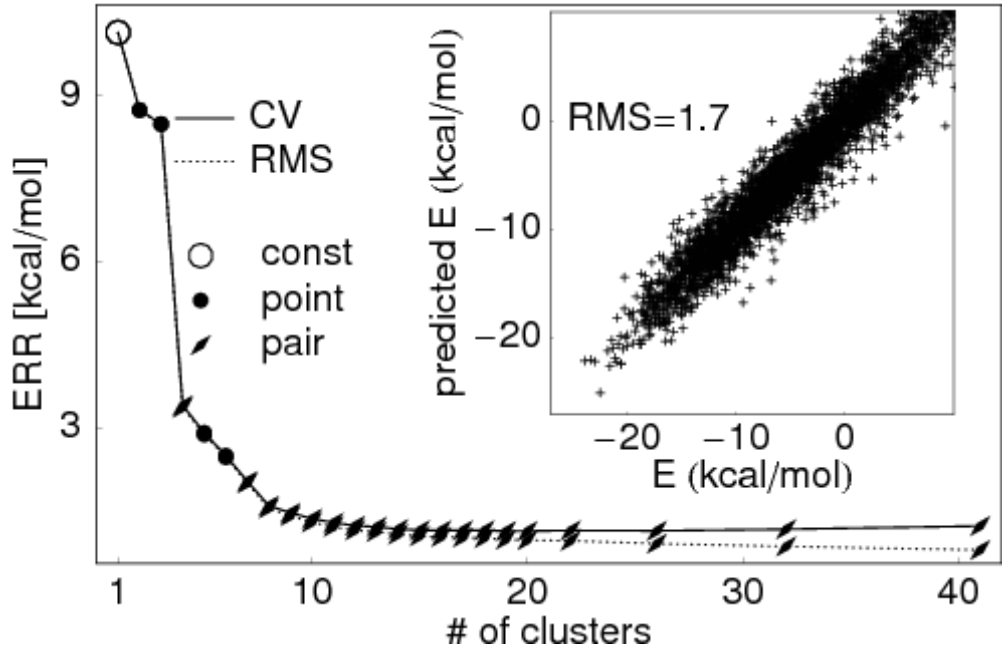}
\caption{RMS and CV scores vs.\ \# of clusters included for
coiled-coil fitting.
Inset: CE predicted vs.\ atomistic $E_{\min}$ for 3995 random
sequences (only $E_{\min} \leq 10$ kcal/mol shown).} \label{fig:cc}
\end{figure}
Our first case study involves the coiled coil, a common and
well-characterized protein interaction interface.  (Fig \ref{fig:peptide-structure}). An
ideal coiled-coil backbone possesses a screw
axis with a repeating unit every 7 residues (a heptad) as well as
C2 symmetry about the coil axis \cite{Crick1953Packing}. We use a
unit cell (highlighted on Fig.\ \ref{fig:peptide-structure}) consisting
of 4 heptads. With the assumption that significant interactions are short range,
the unit cell incorporates all clusters important to
describe coiled-coil stability. Only 4
sites in each heptad are each modeled as one of 16 selected AA
species (the 3 remaining sites are set to Ala). These 4 sites have
been shown, in many cases, to be sufficient to determine
coiled-coil dimerization preferences and other properties
\cite{Oshea1992Mechanism,Fong2004Predicting}. The backbone is
extended by an identical unit cell sequence at both ends to avoid
edge effects. The optimized sidechain configurations correspond to
$E_{\min}$ of the entire protein. The energy of the
central unit cell plus half of its interaction with the rest of
the system is presented.

Our training set consists of 21,066 randomly chosen sequences
weighted by $\max(e^{-(E+26)/120},0.01)$.  Truncating the CE at the pair level is sufficient to accurately reproduce the energetics of the
system. The structural symmetry reduces all 137 clusters up to pairs to 1 constant, 4 point and 36 pair-level
independent cluster (7741 independent $J$s). We are therefore able to include
all of them as candidate clusters in the fitting.
Fig.\ \ref{fig:cc} shows the weighted RMS and CV scores of the
least square fitting versus the number of included clusters
(ordered by $\langle |J| \rangle$). Although the RMS decreases
monotonically as expected, the CV score reaches a minimum at 22
clusters, and fluctuates (mostly increases) slightly afterwards.
We thus come to an ``optimal'' set of 22 clusters (3676 $J$s) for
energy prediction, with weighted RMS = 1.0 kcal/mol and CV = 1.1
kcal/mol. The most significant EIs are found to correspond to
residues that mediate contacts between different helices, in
agreement with biologists' intuition about the system.

To test the predictive character of the CE we compare its energy for 3995 random sequences not included in training to the directly calculated energy ( Fig.\ \ref{fig:cc} inset). The unweighted RMS error is 2.4 kcal/mol for all
energies and 1.7 kcal/mol for -26$< E_{\min} <$10 kcal/mol.
The error is sufficiently small for such applications as sequence
optimization, and is comparable with the accuracy of the
underlying energy model. We trade such a small error for being
able to predict the optimal energy of any sequence by
summation of EIs for 22 clusters, as opposed to performing global
optimization in side-chain conformation space of $5.9\times
10^{55}$ on average. Even compared to the highly efficient DEE
method for sidechain positioning, the time to calculate $E_{\min}$
of a sequence is reduced from $\sim$200 sec to $\sim$1 $\mu$s with
our coarse-grained Hamiltonian, a $2 \times 10^8$-fold
acceleration.

\begin{figure}[tph]
\includegraphics[width=0.8 \linewidth]{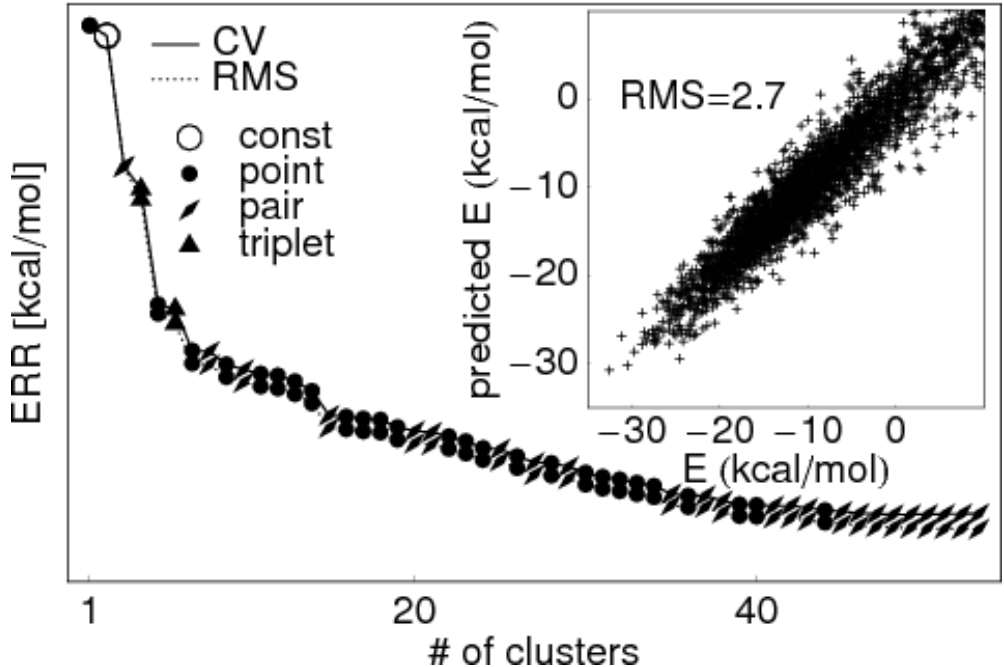}
\caption{RMS and CV scores for Zn-finger fitting.
Inset: CE predicted vs.\ atomistic $E_{\min}$ for 4000 random
sequences (only $E_{\min}\leq 10$
kcal/mol shown). } \label{fig:zf}
\end{figure}
As a second application we consider the zinc-finger, a conserved
DNA-binding fold (Fig \ref{fig:peptide-structure}A). The backbone of Zif268 (PDB ID 1ZAA)
residues 33-60 is used as a model Zn-finger structure. Following Mayo et al.\ \cite{Dahiyat1997De}, we
consider a sequence space in which 2 sites are fixed, 1 site has 7 candidate species,
18 sites have 10 and the other 7 sites have 16.  No symmetry condition is
applied. The training set consists of 29,864 random sequences
weighted by $\max(e^{-(E+35)/100}, 0.01)$. Since there are too many pairs (325
pairs, or $4 \times 10^4$ $J$s) to easily include in one
single fitting, we start with constant and point terms and
add pairs one by one to the existing clusters, retaining a pair
if it decreases CV. We iterate until no new pair can be
selected. However, truncation at pairs leads to an
unsatisfactory fitting with CV$>6$ kcal/mol. Instead of trying
all 2,600 triplets, we use the characteristics of the poorly
fitted sequences ${\cal A}:|\Delta E|>10$ kcal/mol to locate
important triplets. We calculate the information content
$I^i = \ln m^i  - S(p(\sigma^i|{\cal A}))$, $I^{ij}=
\ln m^i m^j  - S(p(\sigma^i \sigma^j|{\cal A})) -I^i -I^j $
for each site $i$ and each pair $\{i,j\}$ out of the AA
distribution in $\cal A$ ($S(p)=-\sum_{\{p\}}p \ln p$ denotes entropy). Four sites have large $I^i$ that are
almost exclusively occupied by aromatic sidechains W, H, Y and F.
Five out of the 6 pairs formed by these sites have significant
$I^{ij}$. Located in proximity to each other, these sites
constitute 2 triplet clusters (see fig.\ \ref{fig:peptide-structure}A).
Thus we use one constant, 26 point, 24 pair, and 2 triplet
clusters (5692 $J$s in total) for fitting. RMS and CV scores
versus the number of clusters included are shown in Fig.\
\ref{fig:zf}. The two triplets 
are found to be indispensable in correctly reproducing the
energies. This demonstrates the existence of complex correlation
in a globular protein, and the CE provides a systematic,
quantitative way of identifying such correlated sites. Prediction of
4000 random Zn-finger sequences is shown in Fig.\ \ref{fig:zf} inset.
Again a reasonably good accuracy of 2.7 kcal/mol
 for $-35 < E_{\min}<10$ kcal/mol is obtained. Although a larger prediction error 15.4 kcal/mol is obtained with all energies, high energy sequences are correctly detected. Such error is traded for a remarkable reduction in
search space: from $1.4\times 10^{60}$ to $1.9 \times 10^{27}$
states.

In summary, we have demonstrated how the energetics of a protein
with pre-defined backbone can be coarse-grained to a function of
sequence only.
The expansion's accuracy can be systematically improved. We have successfully applied the method to two distinct
families of proteins, and found two different types of interactions determining stability.
The accuracy of the CE predictions implies that this much simpler
expression can be used in place of traditional Hamiltonians,
dramatically improving computational efficiency.

The CE methodology can be coupled with any energy model, e.g. more accurate Hamiltonians or experimentally determined energies, and properties other than energy are potentially expandable. Thus, it can be extended to treat any multi-species search problem for which an appropriate scoring scheme can be generated. In structural biology, this includes modeling not only protein stability, but protein interaction specificity, DNA and RNA structure, protein-DNA interactions, and potentially the interactions of small-molecule pharmaceuticals. We are optimistic
that the method will find a wide range of practical applications in biology
research.

This work is supported by funding from the DuPont-MIT Alliance to
GC and NIH grant GM67681 to AK. FZ thanks M.\ Kardar for critical
reading of the manuscript.


\end{document}